\documentclass[sigconf]{acmart}

\usepackage{booktabs} 
\usepackage{subfigure}
\usepackage{amsmath}
\usepackage{color}
\usepackage{amsmath}
\usepackage{amssymb}
\usepackage{mathrsfs}
\usepackage[normalem]{ulem}
\usepackage{geometry}
\usepackage{enumitem}
 \usepackage{multirow}
\usepackage[ruled,vlined]{algorithm2e}
\usepackage{algorithm2e}

\usepackage{ulem}
\usepackage[nomargin,inline,marginclue,draft]{fixme}
\usepackage{balance}
\usepackage{diagbox}
\usepackage{changepage}
\usepackage{multirow}

\newlength\savedwidth

\copyrightyear{2018} 
\acmYear{2018} 
\setcopyright{acmlicensed}
\acmConference[CIKM '18]{The 27th ACM International Conference on Information and Knowledge Management}{October 22--26, 2018}{Torino, Italy}
\acmBooktitle{The 27th ACM International Conference on Information and Knowledge Management (CIKM '18), October 22--26, 2018, Torino, Italy}
\acmPrice{15.00}
\acmDOI{10.1145/3269206.3269234}
\acmISBN{978-1-4503-6014-2/18/10}

\newcommand{\para}[1]{{\vspace{1pt} \bf \noindent #1 \hspace{10pt}}}

\usepackage{caption}
\begin{document}
\title{Recommender Systems with Characterized Social Regularization}

\author{Tzu-Heng Lin, Chen Gao, Yong Li}
\affiliation{%
  \institution{Beijing National Research Center for Information Science and Technology\\Department of Electronic Engineering\\Tsinghua University}
}
\email{lzhbrian@gmail.com, gc16@mails.tsinghua.edu.cn, liyong07@tsinghua.edu.cn}

\begin{abstract}
Social recommendation, which utilizes social relations to enhance recommender systems, has been gaining increasing attention recently with the rapid development of online social network.
Existing social recommendation methods are based on the fact that users preference or decision is influenced by their social friends' behaviors. However, they assume that the influences of social relation are always the same,
which violates the fact that users are likely to share preference on diverse products with different friends. In this paper, we present a novel CSR (short for \textbf{C}haracterized \textbf{S}ocial \textbf{R}egularization) model by designing a universal regularization term for modeling variable social influence. Our proposed model can be applied to both explicit and implicit iteration. 
Extensive experiments on a real-world dataset demonstrate that CSR significantly outperforms state-of-the-art social recommendation methods.
\end{abstract}


\maketitle

\vspace{-0.2cm}
\section{Introduction}\label{sec:Intro}
\vspace{-0.1cm}

Traditional recommender systems that learn user's preference from single domain's data often suffer from data sparsity and cold start problem. To overcome them, social recommendation utilizes user's social relations as auxiliary information to aid learning preference. It relies on the fact that users preference is similar or influenced by their social friends.
Over the past recent years, extensive researches have been working on how to utilize social information to improve recommendation performance \cite{socialrecommendationreview, socialmf, soreg, socialmfbpr, soregbpr, sorec, ste, mtrust, locabal}.

Despite their effectiveness, we argue that all the existing social recommendation models suffer from a limitation:
the social relation between users influences their preference on all items, which violates the fact that a user shares preference on an item with only a part of friends in real life. For example, a user may have similar preferences on book with her schoolmates or workmates, while have similar tastes on food with her family members.

To solve this shortcoming of all existing methods for social recommendation, we propose a Characterized Social Regularization (CSR) model that considers user's various similarities with different friends. Briefly, our method models the item-specific preference similarity of friends to efficiently and accurately learn the social influence on preferences. Specifically, we design a general item-specific regularization term in objective function, which can be easily adapted to most existing recommendation models. 
Through this way, we address the limitation of existing methods, and make models more effective and expressive in social recommendation.

In summary, the contributions of this paper are as follows:
\begin{itemize}[leftmargin=*]
    \item To the best of our knowledge, we are the first to model the characteristics of social relations in the field of social recommendation.
    \item We introduce a novel CSR method that effectively models the characteristics of social relations' influence, and prove that existing social recommendation models can be interpreted as a simplest special case of our CSR model.
    \item Extensive experiments on a real-world dataset show that our method CSR outperforms the state-of-the-art solutions.
\end{itemize}

\vspace{-0.2cm}
\section{PRELIMINARIES}\label{sec:pre}
\vspace{-0.1cm}

We first formulate the task of social recommendation, and then introduce the existing social regularization models by highlighting the limitations to motivate our work.

\vspace{-0.2cm}
\subsection{Problem Formation} \label{subsec:problem-formation}
\vspace{-0.1cm}
In social recommendation, social relations between users are considered as auxiliary information to help learning users' preferences. The target is the same with traditional recommendation tasks, \textit{i.e.}, estimating the unobserved values of the interaction matrix $\mathbf{\mathbf{Y}_{\textit{M}\times{\textit{N}}}} = \{{y}_{ui}\}$ with $M$ and $N$ denoting the number of users and items. 
Specifically, for explicit interactions, ${y}_{ui}$ of user $u$ and item $i$ is a continuous score; while for implicit interactions, it is a binary value. We further use $\mathbf{U}$ and $\mathbf{V}$ to denote set of user and item, respectively, and denote the social relations between users as $\mathbf{S}$ = $\{\mathbf{s}_{uu'}|u,u'\in{\mathbf{U}}\}$. It is a vector representing various relationship between two users $u$ and $u'$, such as friends, following, followed, and etc. In the existing works, this vector is 1-dimension, which is a constant value 1 or a continuous value between 0 and 1 standing for social strength. Then, the task of social recommendation is to estimate the unobserved values in interaction matrix $\bf{Y}$ with the help of social relation $\mathbf{S}$.

\vspace{-0.2cm}
\subsection{Social Regularization} \label{subsec:current-models}
\vspace{-0.1cm}
Regularization, as a widely used technique in matrix factorization approach, sets some limitations to latent vectors through a added regularization term in objective function.
Social regularization is a regularization term considering social relation $\mathbf{S}$.
We denote the added term in social regularization as $Social(\mathbf{\Omega}, \mathbf{S})$, where $\mathbf{\Omega}$ denotes parameters of latent model.
Now, we discuss two mainstream methods utilizing social regularization.

    \para{Distance to weighted sum of friends~\cite{socialmf, socialmfbpr}.}
    This method aims to minimize the distance of a user's latent vector with the weighted sum of the connected users' vector as possible. The added social regularization term is expressed as follows:
    \begin{equation} \label{eqn:loss-socialmf}
        Social(\mathbf{\Omega}, \mathbf{S}) = \sum_{u \in \mathbf{U}} ||\mathbf{p}_u - \sum_{{s}_{uu'} \in \mathbf{S}} \mathbf{s}_{uu'} \mathbf{p}_{u'} ||_2^2,
        \vspace{-0.2cm}
    \end{equation}
    where ${s}_{uu'}$ is a continuous value between 0 and 1 standing for social strength and $\mathbf{p}$ denotes users' latent vector. This method is used in \cite{socialmf} for explicit interactions and in \cite{socialmfbpr} for implicit interactions, respectively.
    
    \para{Sum of weighted distance to friends~\cite{soreg, soregbpr}.}
    This method aims at minimizing the sum of weighted distance between latent vector of a user and her connected users' vector, in which the weight depends on the social strength. The added social regularization term is expressed as follows,
    \begin{equation} \label{eqn:loss-soreg}
            Social(\mathbf{\Omega}, \mathbf{S}) = \sum_{u \in \mathbf{U}} \sum_{{s}_{uu'} \in \mathbf{S}} {s}_{uu'} || \mathbf{p}_u - \mathbf{p}_{u'} ||_2^{2},
            \vspace{-0.2cm}
    \end{equation}
    where symbols are consistent with (\ref{eqn:loss-socialmf}). This method is used in \cite{soreg} for explicit interactions and in \cite{soregbpr} for implicit interactions, respectively.

 As discussed earlier in the Introduction, all these existing methods assume $\mathbf{s}$ is constant for a given pair $(u, u')$, ignoring the fact that the connection $\mathbf{s}$ may vary for interactions with different products. Specifically, preference of $u$ and $u'$ may be very similar on some products, while may be totally different on other products. 
 As a result, social relations' influences on users' preference should be variable on diverse products.
 To address the shortcoming of existing methods, our proposed solution takes the variable influences into account by modeling different impact on diverse products.
 
\vspace{-0.2cm}
\section{Our CSR Solution}\label{sec:model}
\vspace{-0.1cm}

\subsection{Learning from Interactions}
\vspace{-0.1cm}
Since social relation is regarded as side information to help improving recommendation, competitive collaborative filtering model to learn from historical interactions is a basic solution. Here, we rely on Matrix Factorization (MF), a widely used latent factor model in recommender system. 
Let $\mathbf{P}\in{\mathbb{R}^{K\times{M}}}$ and  $\mathbf{Q}\in{\mathbb{R}^{K\times{N}}}$ be latent user and item features matrices, with column vectors $\mathbf{p}_u$ and $\mathbf{q}_i$ representing $K$-dimensional user-specific and item-specific latent feature vectors of users $u$ and item $i$, respectively. 
Since the target of social recommendation is to estimate the unobserved values in interaction matrix $\mathbf{Y}$, MF model tries to decompose $\mathbf{Y}$ as $\mathbf{Y} = \mathbf{P}^\text{T}\mathbf{Q}$.
Then the objective function of MF model can be denoted as:
\begin{equation}\label{eqn:loss-rating}
    \min_{\mathbf{P}, \mathbf{Q}}\mathcal{L}(\mathbf{Y}, \mathbf{P}, \mathbf{Q}) = \sum_{u=1}^{M}\sum_{i=1}^{N} (\mathbf{Y}_{ui} - \mathbf{p}_u^T\mathbf{q}_i)^{2},
\end{equation}
where $L_2$ regularization terms of latent matrices $\mathbf{P}$ and $\mathbf{Q}$ are omitted for simplification.

\vspace{-0.2cm}
\subsection{Characterized Social Regularization}
\vspace{-0.1cm}
As introduced in Section~\ref{subsec:current-models}, regularization is extensively used in social recommendation. In our model, we also utilize regularization technique to constrain the learning of MF model. The motivation of our CSR model is to use social relations to constrain learning from interactions, which is the same with existing methods~\cite{socialmf, socialmfbpr, soreg, soregbpr}. Then, the objective function of MF models considering social regularization can be denoted as:
\begin{equation}
    \min_{\mathbf{P}, \mathbf{Q}}\mathcal{L}(\mathbf{Y}, \mathbf{P}, \mathbf{Q}, \mathbf{S}) = 
    \sum_{u=1}^{M}\sum_{i=1}^{N} (\mathbf{Y}_{ui} - \mathbf{p}_u^T\mathbf{q}_i)^{2} 
    + \lambda_{s} Social(\mathbf{P}, \mathbf{Q}, \mathbf{S})
\end{equation}
where $\lambda_{s}$ is a positive hyper-parameter controlling the weight of social information compared with interaction data.

\para{Dimension-weighted Distance.}
In latent factor model for MF, embedding vector for a user encodes his/her latent interest and embedding vector for an item encodes its features. Considering social influence, a user's interest is similar with his/her friends, which means the embedding vector of user is similar with that of friends. This motivates the two mainstream social regularization methods introduced in Section~\ref{subsec:current-models}, which simply utilize Euclidean distance to build $Social(\mathbf{P}, \mathbf{Q}, \mathbf{S})$ to limit similarity.

However, from the perspective of representation learning,  $\mathbf{q}_{ik}$, the $k$-th dimension in the $K$-dimensional item latent vector, may represent $k$-th feature of item $i$; similarly $\mathbf{p}_{uk}$ may users' preference on the $k$-th item feature. As mentioned in the Introduction, a user may share different preference on various products with different friends. 
Then we can assume two users have similar preference on an item $i$ is due to some dimensions' of item features. In other words, embedding vector of this item has a larger value in those dimensions.
Therefore, we design a dimension-weighted distance rather than a simple Euclidean distance to guarantee the similarity of social-connected users.
For $k$-th dimension of relationship between user $u$ and one of his/her friends $u'$, we use ${s}^{k}_{uu'}$ to denote the $k$-th weight, then we obtain a weight vector: $\mathbf{s}_{uu'}$. Based on this vector, the characterized regularization term can be denoted as:
\begin{equation} \label{loss:csr-general}
Social(\mathbf{P}, \mathbf{Q}, \mathbf{S}) =                    
    \sum_{(u, u') \in \mathbf{S}} || (\mathbf{p}_u - \mathbf{p}_u') \circ \mathbf{s}_{uu'} ||_2^{2},
\end{equation}
where $\circ$ means a Hadamard (element-wise) product.

\para{Product-sharing Based Social Relations}
Although the designed social relations, weight vector $\mathbf{s_{uu'}}$, can help modeling social influence theoretically, 
it is not possible for them to be learned automatically, directly optimizing the above objective function will result in a trivial all zero solution for $\mathbf{s_{uu'}}$.
In practice, users exchange opinions on various kinds of products, such as movies, news, etc., through sharing with friends.
Therefore, to some extent, a sharing behavior or discussion on item $i$ between user $u$ and a friend $u'$ is a social relation accompanied by item $i$. Thus, we can infer that $u$ and $u'$ share tastes on item $i$ and those items similar with $i$.
Since the embedding vector of shared item $\mathbf{q}_i$ represents item features, it can be straightly used as weight vector $\mathbf{s_{uu'}}$. 
The weight vector can also be built from other data or information of social domain certainly. This paper only focuses on the general framework, and leaves the specific design as future work. 

Thus, we formulate our proposed social regularization term as:
\begin{equation}
    Social(\mathbf{P}, \mathbf{Q}, \mathbf{S}) =                    
    \sum_{(u, u', i) \in \mathcal{D}} || (\mathbf{p}_u - \mathbf{p}_u') \circ \mathbf{q}_{i} ||_2^{2},
\end{equation}
where $\mathcal{D}$ is the triplets of product-sharing logs, which is utilized to build social relations $\mathbf{S}$, and ${(u, u', i) \in \mathcal{D}}$ means user $u$ shows preference on item $i$ with his/her friend $u'$.
Then the objective function of our CSR model is:
\begin{equation}
    \begin{split}
    \min_{\mathbf{P}, \mathbf{Q}}\mathcal{L}(\mathbf{Y}, \mathbf{P}, \mathbf{Q}, \mathbf{S}) = 
    & \sum_{u=1}^{M}\sum_{i=1}^{N} (\mathbf{Y}_{ui} - \mathbf{p}_u^T\mathbf{q}_i)^{2} \\
    & + \lambda_{s} \sum_{(u, u', i) \in \mathcal{D}} || (\mathbf{p}_u - \mathbf{p}_u') \circ \mathbf{q}_{i} ||_2^{2}.
    \end{split}
\end{equation}

\begin{figure}[t]
    \begin{center}
    \mbox{
        \subfigure[Existing models]{\includegraphics[width= 3cm]{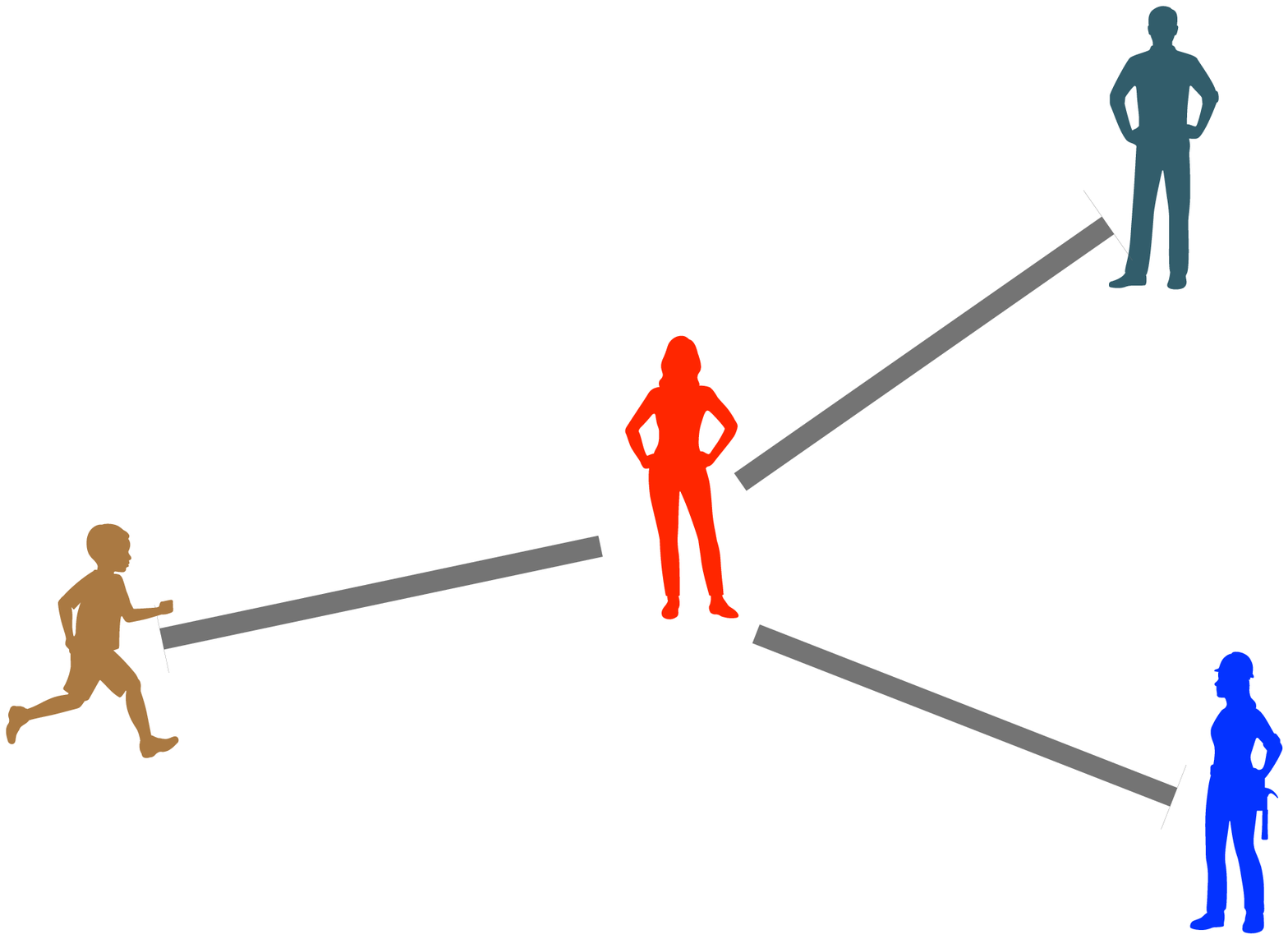}}
        \hspace{0.5cm}
        \subfigure[Our CSR model]{\includegraphics[width=3cm]{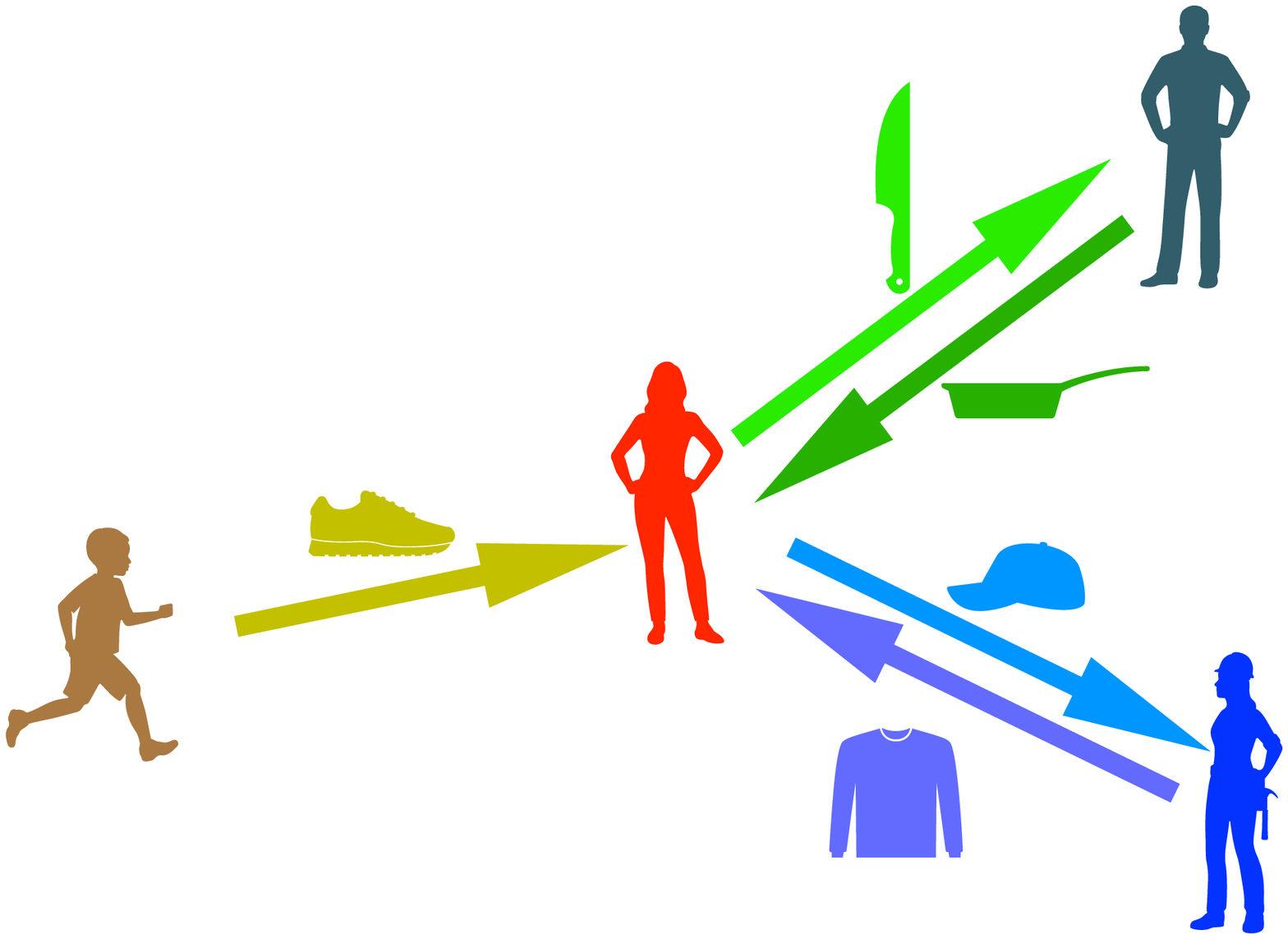}}
    }
    \end{center}
    \caption{Comparison of existing models and our CSR model}\label{Fig:Comparison}
\end{figure}

\vspace{-0.2cm}
\subsection{Training}\label{subsec:training}
\vspace{-0.1cm}
    We optimize parameters with stochastic gradient descent (SGD), and implement it on Tensorflow\footnote{https://www.tensorflow.org/}, which provides the function of automatic differentiation. 
    So here we omit the derivation of the objective function.
    We introduce our model in a point-wise manner for explicit interaction data, and for implicit interaction data, in fact, our CSR model can also be trained in a pair-wise manner~\cite{bpr}.
    Therefore, our proposed model can be adapted to these two kinds of interaction data just by switching the training manner.

\vspace{-0.2cm}
\subsection{Discussions}
\vspace{-0.1cm}
    Here, we summary some desirable properties of CSR.
    Firstly, it is obviously that if we set the value of every dimension of $\mathbf{s}$ to be 1 in (\ref{loss:csr-general}), then our proposed CSR model is transferred to the methods discussed in Section~\ref{subsec:current-models}. Thus, existing social regularization methods can be taken as a special case of our CSR model.
    Secondly, the regularization term in our CSR model can be adapted to the objective function of most social recommendation models, not only for the MF model, which means the model learning from historical interaction can be diverse. Furthermore, this makes our CSR model able to handle complicated tasks and various interaction data.

\vspace{-0.2cm}
\section{Experiments}\label{sec:experiments}
\vspace{-0.1cm}

In this section, we conducted extensive experiments on a real-world dataset to answer the following three research questions:
\begin{itemize}[leftmargin=*]
\item \textbf{RQ1:} How does our proposed CSR model perform as compared with state-of-the-art social recommendation methods?
\item \textbf{RQ2:} How do the key hyper-parameters affect performance of our CSR model? 
\item \textbf{RQ3:} How does number of users' social relations affect recommendation performance of our CSR model?
\end{itemize}

\vspace{-0.3cm}
\subsection{Experimental Settings}
\vspace{-0.1cm}

    \para{Dataset}
        We experimented with a dataset collected from Beibei (https://www.beibei.com), which is the largest E-commerce platform for maternal and infant products in China. This dataset includes user logs from 2017/06/01 to 2017/06/30. 
        This dataset is made up of two parts: purchase record containing users' purchase log and sharing record containing users' item-sharing logs. The statistics of the dataset are summarized in Table~\ref{tab:Dataset}.
        Note that, here we learn CSR model based on pair-wise loss~\cite{bpr} for better performance.
        Due to the need for sharing data, no publicly available dataset are suitable for our methods. An interesting future work would be discovering more possible approaches to build explicit social relations.
        
        \begin{table}[t]
            \begin{center}
            \small
                \caption{Statistics of our utilized Beibei dataset}\label{tab:Dataset}
                \begin{tabular}{|c|c|c|c|}
                    \hline
                    {\bf \# Users} & {\bf \# Items} & {\bf \# Purchases} & {\bf \# Shares} \\\hline
                    337  & 553 & 2,572 & 476 \\\hline
                \end{tabular}
            \end{center}
        \end{table}
    \para{Evaluation Metrics}
        To evaluate the performance, we adopted the \emph{leave-one-out} evaluation method with the following metrics.
        \begin{itemize}[leftmargin=*]
            \vspace{-0.1cm}
            \item \textbf{HR@K:} Hit Rate
            \item \textbf{NDCG@K:} Normalized Discounted Cumulative Gain
            \vspace{-0.1cm}
        \end{itemize}
    \para{Baselines} We compared our CSR model with six baseline methods can be used in our implicit dataset.
    The first group contains methods only utilizing interaction data.
        \begin{itemize}[leftmargin=*]
            \vspace{-0.1cm}
            \item \textbf{Random (Rand)} This method randomly orders the items.
            \item \textbf{ItemPop} This method ranks items base on their popularity, as judged by the number of historical interactions.
            \item \textbf{BPR \cite{bpr}} This is a basic method based on pairwise learning.
            \vspace{-0.1cm}
        \end{itemize}
    The second group contains social recommendation methods.
        \begin{itemize}[leftmargin=*]
            \vspace{-0.1cm}
            \item \textbf{SocialBPR \cite{socialmfbpr}} This method optimizes a pairwise loss with a distance-to-weighted-sum-of-friends social regularization term. 
            \item \textbf{UGPMF \cite{soregbpr}} This method optimizes a pairwise loss with a sum-of-weighted-distance-to-friends social regularization term.
            \item \textbf{SBPR \cite{sbpr}} This method assumes that users tend to assign higher ranks to items that their friends prefer.
            \vspace{-0.1cm}
        \end{itemize}

    \para{Parameter Settings}
        Following \cite{sorec, soreg, socialmf}, we set the weights of regularization terms $\lambda_P$ and $\lambda_Q$ for $P$ and $Q$ to a trivial value 0.01.
        To make the experiments reasonable, we also use grid search on learning rate $\alpha$, dimensionality $K$ and the weight of social regularization term $\lambda_s$ to find the best performance for each models.
        In the following sections, we report the best parameter settings for each model.

\vspace{-0.2cm}
\subsection{Performance Comparison (RQ1)}
\vspace{-0.1cm}

The experiment results are shown in Table~\ref{tab:Comparision}.
Following findings are observed.
\begin{itemize}[leftmargin=*]
    \vspace{-0.1cm}
    \item Rand and ItemPop achieve the worst performance, which indicates the power of MF model in modeling personalized behaviors.
    \item Generally, models with social information (SBPR, SocialBPR, CSR) outperforms the one without it (BPR) in terms of HR@10, NDCG@10.
    However, due to the sparsity of our social data, some methods (e.g. UGPMF) fail to utilize social information well.
    \item We can observe that CSR significantly and consistently outperforms the best baseline methods in terms of HR@5, NDCG@5, HR@10, NDCG@10
    with a relative improvement of 11.19\%, 16.21\%, 2.85\%, 8.63\%, respectively. This justifies the expressiveness and effectiveness of our model.
    \vspace{-0.1cm}
\end{itemize}

\begin{table}[t]
	\small
	\begin{center}
	\small
    		\caption{Performance comparison}\label{tab:Comparision}
    		\begin{tabular}{|c|c|c|c|c|c|}
    			\hline
    			Method                       & HR@5            & NDCG@5          & HR@10           & NDCG@10         \\\hline          
    			Rand                         & 0.0119          & 0.0082          & 0.0297          & 0.0143          \\\hline     
    			ItemPop                      & 0.0623          & 0.0353          & 0.0920          & 0.0450          \\\hline   
    			BPR \cite{bpr}               & 0.0742          & 0.0401          & 0.1246          & 0.0556          \\\hline         
    			SBPR \cite{sbpr}             & 0.0677          & 0.0357          & 0.1294          & 0.0562          \\\hline      
    			UGPMF \cite{soregbpr}        & 0.0682          & 0.0371          & 0.1231          & 0.0537          \\\hline     
    			SocialBPR \cite{socialmfbpr} & 0.0742          & 0.0397          & 0.1403          & 0.0626          \\\hline    
    			\textbf{Our CSR}                 & \textbf{0.0825} & \textbf{0.0466} & \textbf{0.1443} & \textbf{0.0680} \\\hline
    		\end{tabular}
	\end{center}
\end{table}

\vspace{-0.3cm}
\subsection{Parameter Study of CSR (RQ2)}
\begin{figure}[t]
            \begin{center}
            \mbox{
                \subfigure{\includegraphics[width=3.5cm]{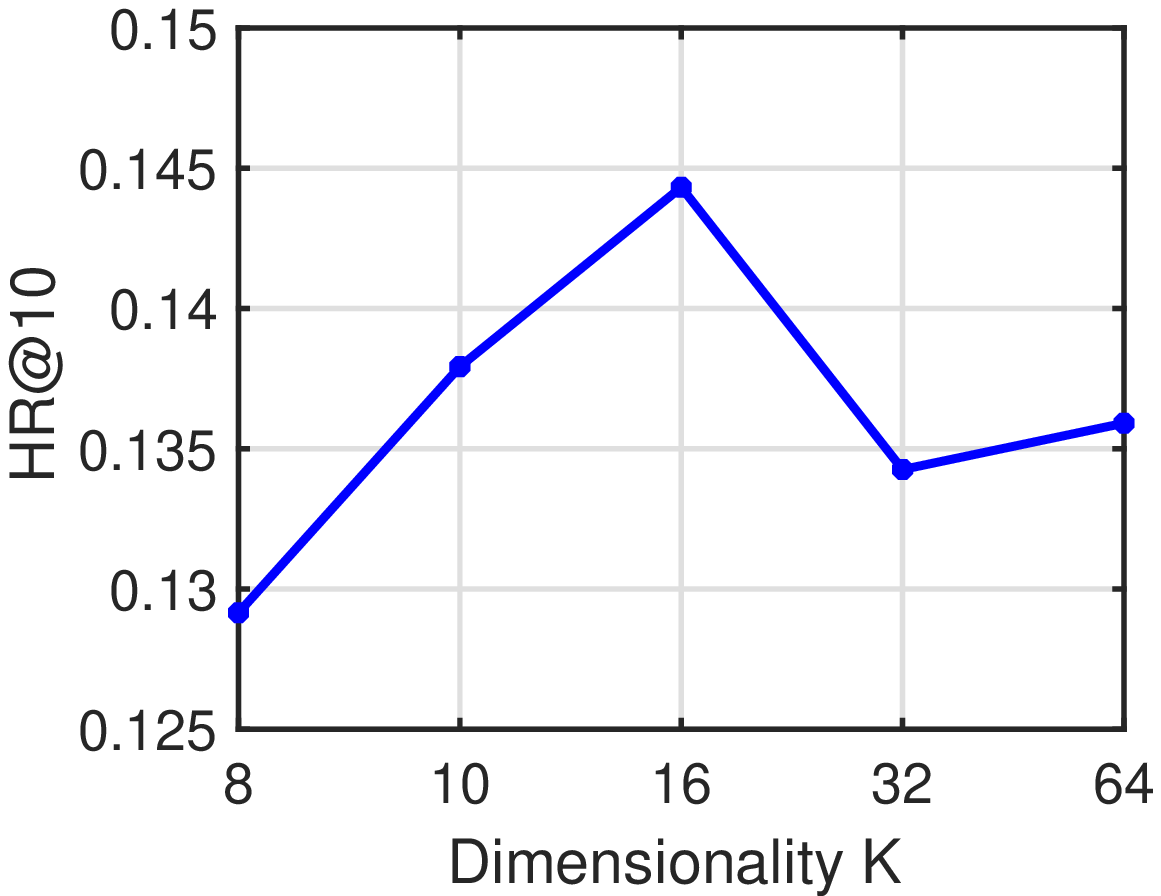}}
                \subfigure{\includegraphics[width=3.5cm]{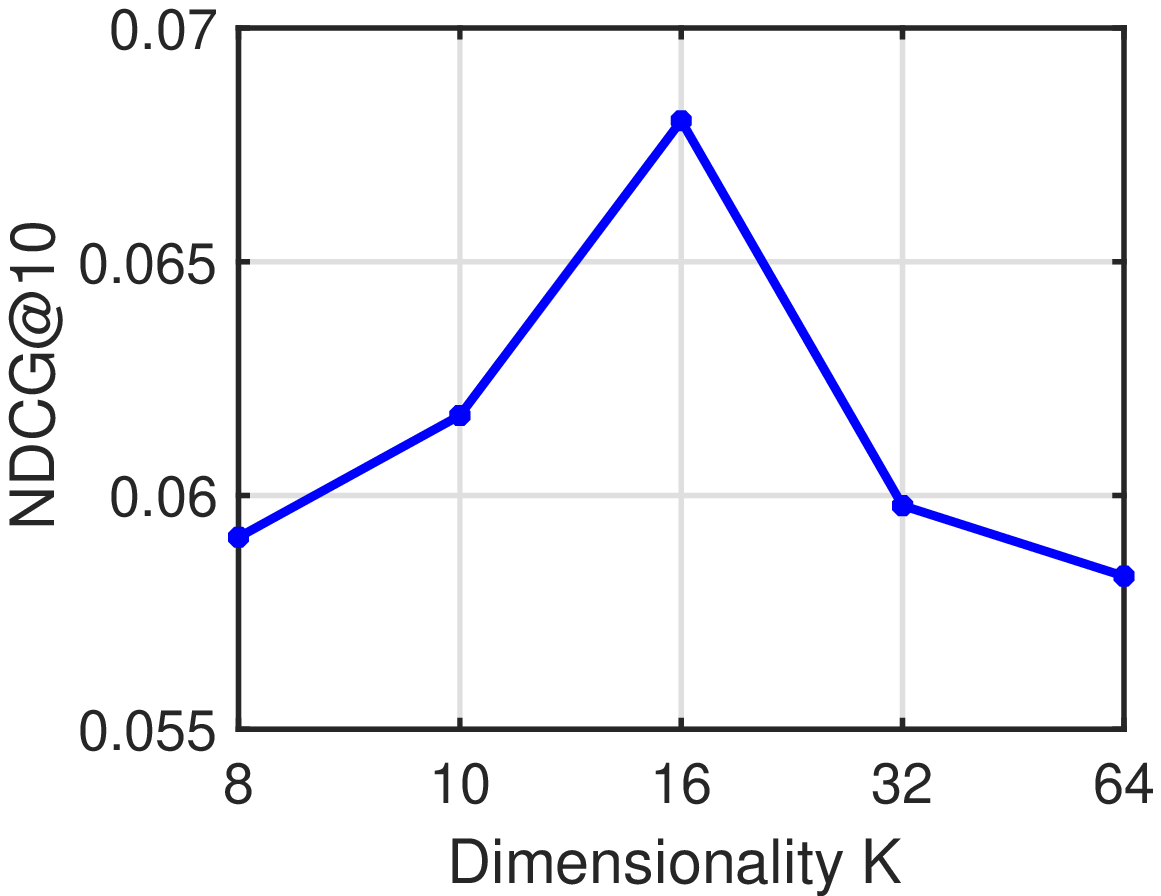}}
            }
            \end{center}
            \caption{Performance with different dimensionality $K$}\label{Fig:RQ2-dim}
            \begin{center}
            \mbox{
                \subfigure{\includegraphics[width=3.5cm]{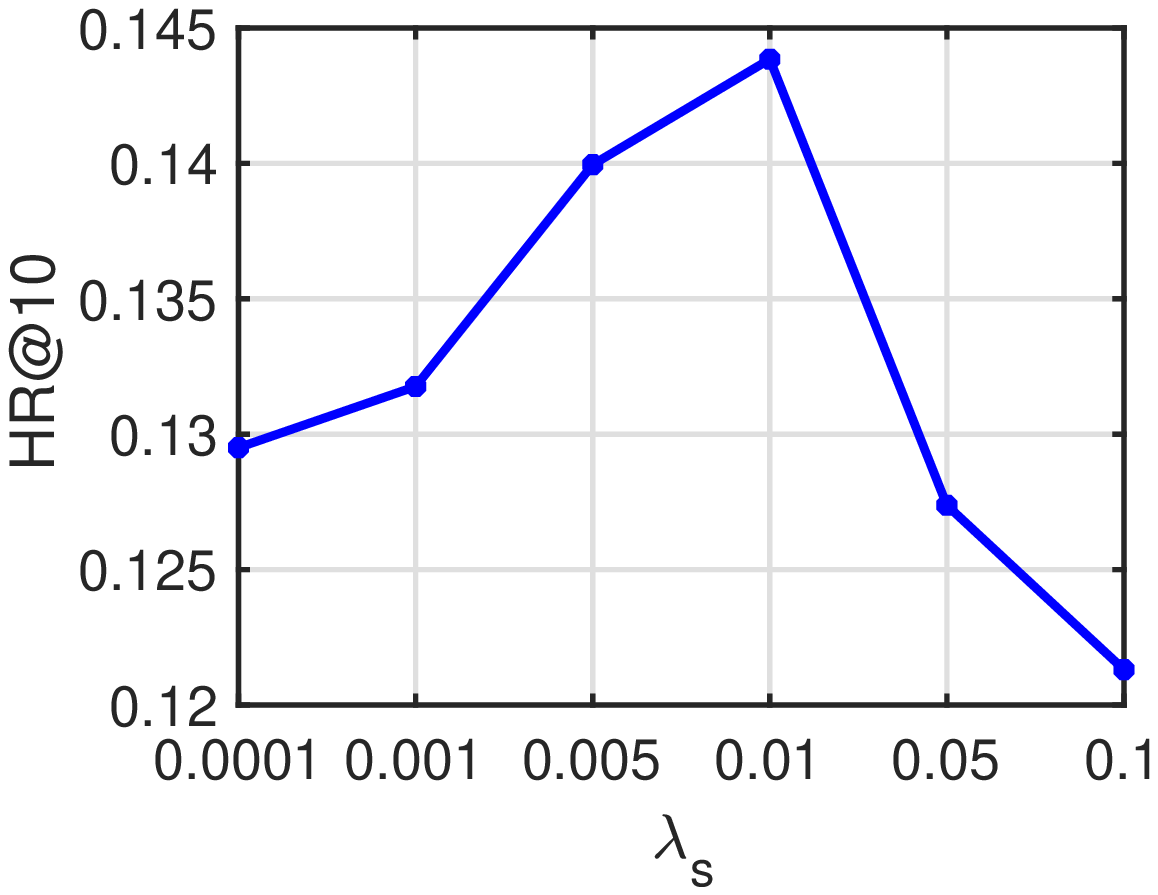}}
                \subfigure{\includegraphics[width=3.5cm]{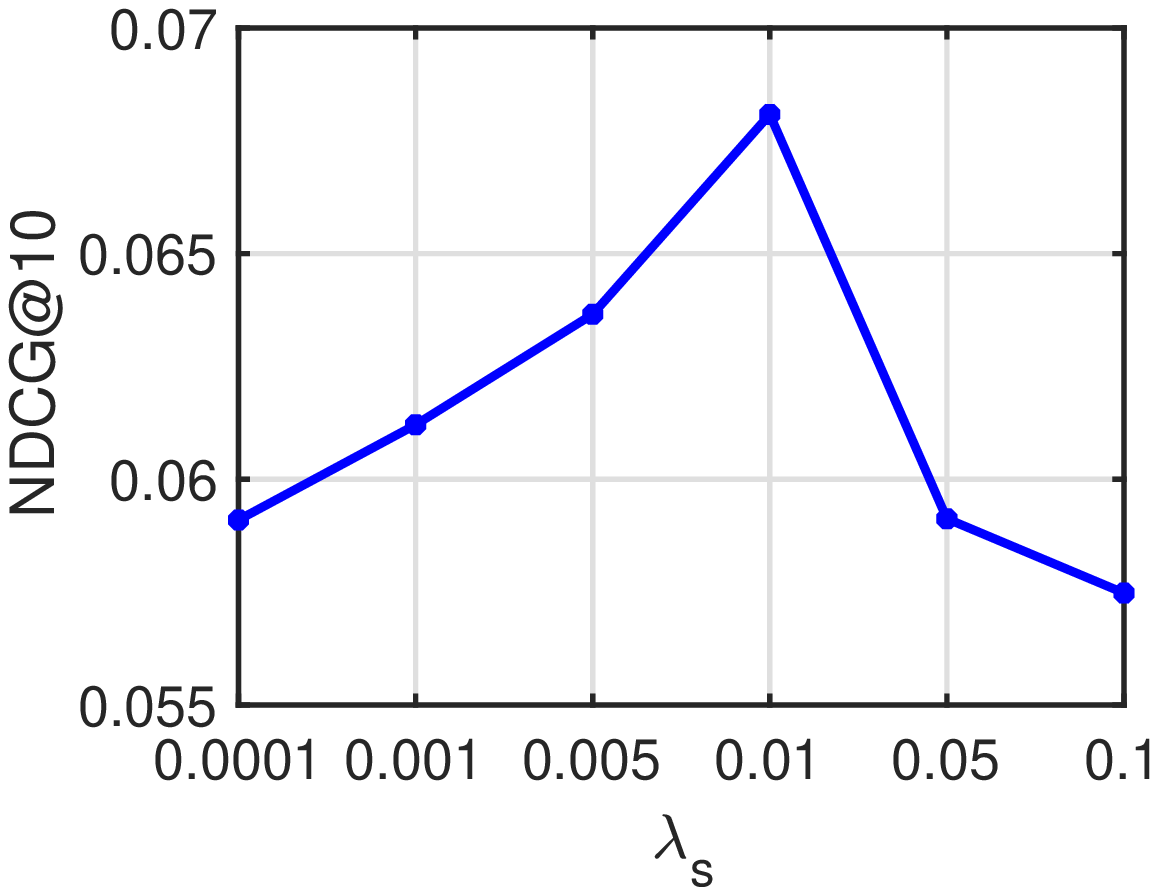}}
            }
            \end{center}
            \caption{Performance with different weight $\lambda_{s}$}\label{Fig:RQ2-lambdas}
        \end{figure}

\vspace{-0.1cm}
    \para{Impact of Dimensionality $K$}
        Intuitively, increasing the dimensionality of latent feature vectors can make models more expressive. However, on the other hand, training models with a larger $K$ needs more data. 
        As seen in Figure~\ref{Fig:RQ2-dim}, in our experiment, we find that a middle size $K = 16$ can have the best performance.
    
    \para{Impact of weight $\lambda_{s}$}
        The weight of social regularization term $\lambda_s$ controls the influence of social data compared with the interaction data.
        A larger value of $\lambda_s$ gives more contribution of the social sharing information. However, it may also cause social term dominates the total loss, and do harm to the performance with less utilization of the interaction data.
        We compare the different settings of $\lambda_s$ in Figure~\ref{Fig:RQ2-lambdas} and find the optimal value for $\lambda_s$ is 0.01 in our experiment.

\vspace{-0.4cm}
\subsection{Impact of Social Relation (RQ3)}
\begin{figure}[t]
        \begin{center}
        \mbox{
            \subfigure{\includegraphics[width=3.8cm]{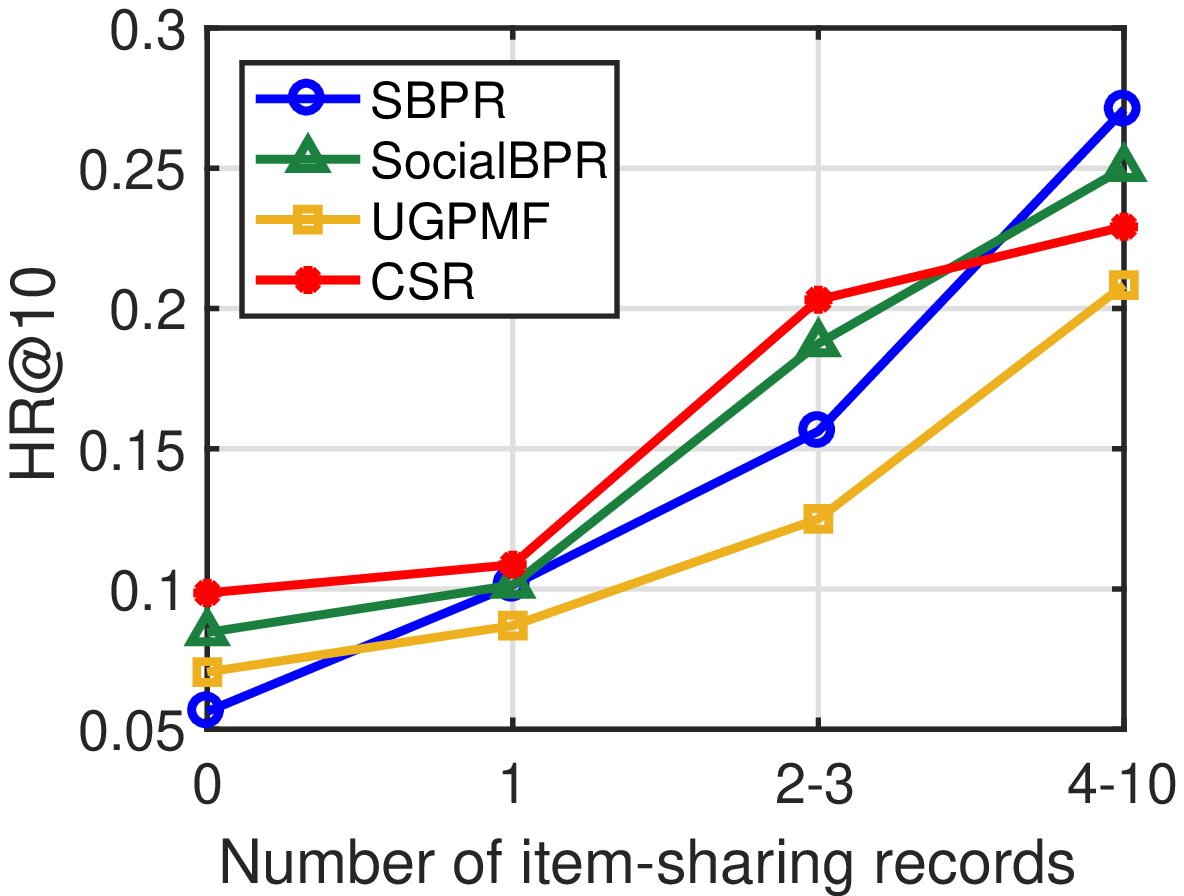}}
            \subfigure{\includegraphics[width=3.8cm]{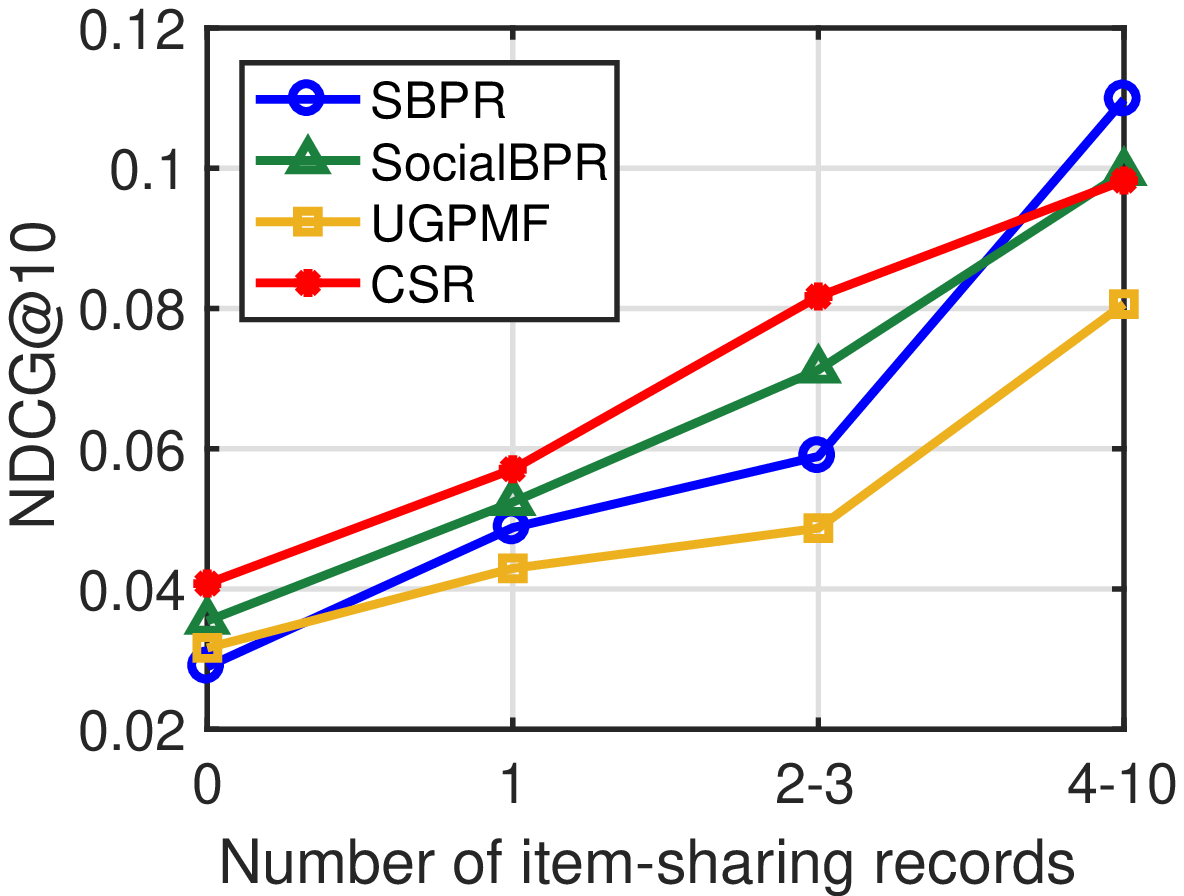}}
        }
        \end{center}
        \caption{Performance on users with different number of social relation}\label{Fig:RQ3}
    \end{figure}

\vspace{-0.1cm}
Since social recommendation models utilize social information to improve recommendation, it is meaningful to study how our model outperforms in making use of social relations.
As users with more relations are sparser, we divide users to four groups according to the number of sharing records: [0, 1, 2-3, $>$3]. This makes sure that the number of users in each group is very close.
The recommendation performances for each group of our CSR and existing methods of social recommendation are shown in Figure~\ref{Fig:RQ3}.

Firstly, we can observe that all methods achieve better recommendation performance for users with more social relations, which demonstrates the importance of social information. 
Secondly, although all models achieve good performance for users with abundant social relations, our CSR model performs best for users with sparse social relations. Even for those users without social relations, our CSR model can improve recommendation via better learned latent user/item vectors.
Therefore, we can conclude that our CSR model can efficiently address the sparsity problem of social data.

\vspace{-0.2cm}
\section{Conclusion}\label{sec:conclusion}
\vspace{-0.1cm}
In this work, we have proposed a novel model \textit{Characterized Social Regularization} (CSR) to generalize the traditional social regularization methods.
We have shown that existing mainstream social recommendation models utilizing regularization techniques can be regarded as special cases of our CSR model.
To the best of our knowledge, this is by far the first approach to address and tackle the problem of modeling the variable characteristics of social relations in the field of social recommendation.
We conducted sufficient experiments on a real-world E-commerce dataset, showing that our proposed CSR outperforms the existing methods.
As future work, we would like to apply CSR to perform recommendation task on explicit interaction data, and we are also interested in exploring more possible approaches to build social relations besides item sharing.

\vspace{-0.2cm}
\begin{acks}
\vspace{-0.1cm}
This work was supported in part by The National Key Research and Development Program of China under grant 2017YFE0112300, the National Nature Science Foundation of China under 61861136003, 61621091 and 61673237, Beijing National Research Center for Information Science and Technology under 20031887521, and research fund of Tsinghua University - Tencent Joint Laboratory for Internet Innovation Technology.
\end{acks}

\vspace{-0.2cm}
\bibliographystyle{ACM-Reference-Format}
\bibliography{bibliography}


\begin{thebibliography}{11}


\ifx \showCODEN    \undefined \def \showCODEN     #1{\unskip}     \fi
\ifx \showDOI      \undefined \def \showDOI       #1{#1}\fi
\ifx \showISBNx    \undefined \def \showISBNx     #1{\unskip}     \fi
\ifx \showISBNxiii \undefined \def \showISBNxiii  #1{\unskip}     \fi
\ifx \showISSN     \undefined \def \showISSN      #1{\unskip}     \fi
\ifx \showLCCN     \undefined \def \showLCCN      #1{\unskip}     \fi
\ifx \shownote     \undefined \def \shownote      #1{#1}          \fi
\ifx \showarticletitle \undefined \def \showarticletitle #1{#1}   \fi
\ifx \showURL      \undefined \def \showURL       {\relax}        \fi
\providecommand\bibfield[2]{#2}
\providecommand\bibinfo[2]{#2}
\providecommand\natexlab[1]{#1}
\providecommand\showeprint[2][]{arXiv:#2}

\bibitem[\protect\citeauthoryear{Du, Li, and Shen}{Du et~al\mbox{.}}{2011}]%
        {soregbpr}
\bibfield{author}{\bibinfo{person}{Liang Du}, \bibinfo{person}{Xuan Li}, {and}
  \bibinfo{person}{Yi-Dong Shen}.} \bibinfo{year}{2011}\natexlab{}.
\newblock \showarticletitle{User graph regularized pairwise matrix
  factorization for item recommendation}. In \bibinfo{booktitle}{{\em ADMA}}.
  \bibinfo{pages}{372--385}.
\newblock


\bibitem[\protect\citeauthoryear{Guo, Ma, Jiang, Chen, and Xing}{Guo
  et~al\mbox{.}}{2015}]%
        {socialmfbpr}
\bibfield{author}{\bibinfo{person}{Lei Guo}, \bibinfo{person}{Jun Ma},
  \bibinfo{person}{Hao-Ran Jiang}, \bibinfo{person}{Zhu-Min Chen}, {and}
  \bibinfo{person}{Chang-Ming Xing}.} \bibinfo{year}{2015}\natexlab{}.
\newblock \showarticletitle{Social trust aware item recommendation for implicit
  feedback}.
\newblock \bibinfo{journal}{{\em JCST\/}} \bibinfo{volume}{30},
  \bibinfo{number}{5} (\bibinfo{year}{2015}), \bibinfo{pages}{1039--1053}.
\newblock


\bibitem[\protect\citeauthoryear{Jamali and Ester}{Jamali and Ester}{2010}]%
        {socialmf}
\bibfield{author}{\bibinfo{person}{Mohsen Jamali} {and} \bibinfo{person}{Martin
  Ester}.} \bibinfo{year}{2010}\natexlab{}.
\newblock \showarticletitle{A matrix factorization technique with trust
  propagation for recommendation in social networks}. In
  \bibinfo{booktitle}{{\em RecSys}}. \bibinfo{pages}{135--142}.
\newblock


\bibitem[\protect\citeauthoryear{Ma, King, and Lyu}{Ma et~al\mbox{.}}{2009}]%
        {ste}
\bibfield{author}{\bibinfo{person}{Hao Ma}, \bibinfo{person}{Irwin King}, {and}
  \bibinfo{person}{Michael~R Lyu}.} \bibinfo{year}{2009}\natexlab{}.
\newblock \showarticletitle{Learning to recommend with social trust ensemble}.
  In \bibinfo{booktitle}{{\em SIGIR}}. \bibinfo{pages}{203--210}.
\newblock


\bibitem[\protect\citeauthoryear{Ma, Yang, Lyu, and King}{Ma
  et~al\mbox{.}}{2008}]%
        {sorec}
\bibfield{author}{\bibinfo{person}{Hao Ma}, \bibinfo{person}{Haixuan Yang},
  \bibinfo{person}{Michael~R Lyu}, {and} \bibinfo{person}{Irwin King}.}
  \bibinfo{year}{2008}\natexlab{}.
\newblock \showarticletitle{Sorec: social recommendation using probabilistic
  matrix factorization}. In \bibinfo{booktitle}{{\em CIKM}}.
  \bibinfo{pages}{931--940}.
\newblock


\bibitem[\protect\citeauthoryear{Ma, Zhou, Liu, Lyu, and King}{Ma
  et~al\mbox{.}}{2011}]%
        {soreg}
\bibfield{author}{\bibinfo{person}{Hao Ma}, \bibinfo{person}{Dengyong Zhou},
  \bibinfo{person}{Chao Liu}, \bibinfo{person}{Michael~R Lyu}, {and}
  \bibinfo{person}{Irwin King}.} \bibinfo{year}{2011}\natexlab{}.
\newblock \showarticletitle{Recommender systems with social regularization}. In
  \bibinfo{booktitle}{{\em WSDM}}. \bibinfo{pages}{287--296}.
\newblock


\bibitem[\protect\citeauthoryear{Rendle, Freudenthaler, Gantner, and
  Schmidt-Thieme}{Rendle et~al\mbox{.}}{2009}]%
        {bpr}
\bibfield{author}{\bibinfo{person}{Steffen Rendle}, \bibinfo{person}{Christoph
  Freudenthaler}, \bibinfo{person}{Zeno Gantner}, {and} \bibinfo{person}{Lars
  Schmidt-Thieme}.} \bibinfo{year}{2009}\natexlab{}.
\newblock \showarticletitle{BPR: Bayesian personalized ranking from implicit
  feedback}. In \bibinfo{booktitle}{{\em UAI}}. \bibinfo{pages}{452--461}.
\newblock


\bibitem[\protect\citeauthoryear{Tang, Gao, Hu, and Liu}{Tang
  et~al\mbox{.}}{2013a}]%
        {locabal}
\bibfield{author}{\bibinfo{person}{Jiliang Tang}, \bibinfo{person}{Huiji Gao},
  \bibinfo{person}{Xia Hu}, {and} \bibinfo{person}{Huan Liu}.}
  \bibinfo{year}{2013}\natexlab{a}.
\newblock \showarticletitle{Exploiting homophily effect for trust prediction}.
  In \bibinfo{booktitle}{{\em WSDM}}. \bibinfo{pages}{53--62}.
\newblock


\bibitem[\protect\citeauthoryear{Tang, Gao, and Liu}{Tang
  et~al\mbox{.}}{2012}]%
        {mtrust}
\bibfield{author}{\bibinfo{person}{Jiliang Tang}, \bibinfo{person}{Huiji Gao},
  {and} \bibinfo{person}{Huan Liu}.} \bibinfo{year}{2012}\natexlab{}.
\newblock \showarticletitle{mTrust: discerning multi-faceted trust in a
  connected world}. In \bibinfo{booktitle}{{\em WSDM}}.
  \bibinfo{pages}{93--102}.
\newblock


\bibitem[\protect\citeauthoryear{Tang, Hu, and Liu}{Tang
  et~al\mbox{.}}{2013b}]%
        {socialrecommendationreview}
\bibfield{author}{\bibinfo{person}{Jiliang Tang}, \bibinfo{person}{Xia Hu},
  {and} \bibinfo{person}{Huan Liu}.} \bibinfo{year}{2013}\natexlab{b}.
\newblock \showarticletitle{Social recommendation: a review}.
\newblock \bibinfo{journal}{{\em SNAM\/}} \bibinfo{volume}{3},
  \bibinfo{number}{4} (\bibinfo{year}{2013}), \bibinfo{pages}{1113--1133}.
\newblock


\bibitem[\protect\citeauthoryear{Zhao, McAuley, and King}{Zhao
  et~al\mbox{.}}{2014}]%
        {sbpr}
\bibfield{author}{\bibinfo{person}{Tong Zhao}, \bibinfo{person}{Julian
  McAuley}, {and} \bibinfo{person}{Irwin King}.}
  \bibinfo{year}{2014}\natexlab{}.
\newblock \showarticletitle{Leveraging social connections to improve
  personalized ranking for collaborative filtering}. In
  \bibinfo{booktitle}{{\em CIKM}}. \bibinfo{pages}{261--270}.
\newblock


\end{thebibliography}

\end{document}